\documentclass[twocolumn,showpacs,amssymb,aps,prl,floatfix]{revtex4}

\usepackage{epsfig}

\newcommand{\bra}{\langle}
\newcommand{\ket}{\rangle}

\newcommand{\Tr}{\hbox{Tr}}

\newcommand{\kv}{{\mathbf k}}

\newcommand{\vecx}{{\mathbf x}}

\newcommand{\vecz}{{\mathbf z}}

\newcommand{\be}{\begin{equation}}
\newcommand{\ee}{\end{equation}}
\newcommand{\bea}{\begin{eqnarray}}
\newcommand{\eea}{\end{eqnarray}}
\newcommand{\bean}{\begin{eqnarray*}}
\newcommand{\eean}{\end{eqnarray*}}

\newcommand{\C}{{\cal C}}
\newcommand{\half} {\frac{1}{2}}
\newcommand{\om} {\omega}

\newcommand{\NNLO}{N$^2$LO}
\newcommand{\NNLOs}{N$^2$LO }

\newcommand{\nn}{\nonumber}

\begin{document}                                                

\title{Effective convergence of the 2PI-$1/N$ expansion for 
nonequilibrium quantum fields}

\author{Gert Aarts}
\author{Nathan Laurie}
\affiliation{Department of Physics, Swansea University, Swansea SA2 8PP, 
United Kingdom}
\author{Anders Tranberg}
\affiliation{Department of Physical Sciences, University of Oulu, P.O.\ 
Box 3000, FI-90014 Oulu, Finland}

\date{September 19, 2008}

\begin{abstract} 

The $1/N$ expansion of the two-particle irreducible effective action 
offers a powerful approach to study quantum field dynamics far from 
equilibrium. We investigate the effective convergence of the $1/N$ 
expansion in the $O(N)$ model by comparing results obtained numerically in 
$1+1$ dimensions at leading, next-to-leading and next-to-next-to-leading 
order in $1/N$ as well as in the weak coupling limit. A comparison in 
classical statistical field theory, where exact numerical results are 
available, is made as well. We focus on early-time dynamics and 
quasi-particle properties far from equilibrium and observe rapid effective 
convergence already for moderate values of $1/N$ or the coupling.

\end{abstract}

\pacs{
}

\maketitle


{\em Introduction --}
 The need to understand the dynamics in heavy ion physics and the quark 
gluon plasma as well as highly nonlinear phenomena in the 
(post-)inflationary universe has lead to substantial development in the 
study of quantum field evolution far from equilibrium. One particular 
approach, firmly based on functional methods in field theory, employs the 
two-particle irreducible (2PI) effective action \cite{Berges:2004yj}. The 
nonperturbative 2PI-$1/N$ expansion \cite{Berges:2001fi,Aarts:2002dj}, 
where $N$ denotes the number of matter field components, has in particular 
proven fruitful.
  In recent years this approach has been applied to a variety of 
nonequilibrium problems in $3+1$ dimensions, related to inflationary 
preheating \cite{Berges:2002cz,Arrizabalaga:2004iw}, 
effective prethermalization \cite{Berges:2004ce}, 
fermion dynamics \cite{Berges:2002wr},
transport coefficients \cite{Aarts:2003bk}
and kinetic theory \cite{Juchem:2004cs,Berges:2005md}, 
slow-roll dynamics \cite{Aarts:2007qu}, 
topological defects \cite{Rajantie:2006gy},
nonthermal fixed points \cite{Berges:2008wm}, 
expanding backgrounds \cite{Tranberg:2008ae},
nonrelativistic cold atoms \cite{Gasenzer:2005ze}, etc.
 Besides these applications, theoretical aspects are under continuous 
investigation: renormalization in equilibrium is by now well understood 
\cite{vanHees:2001ik,Blaizot:2003br,Berges:2004hn}
and practical implementations of renormalization out of equilibrium are 
being developed \cite{Borsanyi:2008ar}. Moreover, progress in adapting 
these methods to gauge 
theories is steady 
\cite{Arrizabalaga:2002hn,Reinosa:2006cm,Borsanyi:2007bf}.

Applications of 2PI effective action techniques are based on truncations, 
employing either a weak coupling or a large $N$ expansion. So far, 
truncations stop at relatively low order: as far as we are aware all 
studies in field theory use next-to-leading order (NLO) truncations in the 
coupling or $1/N$ \footnote{Mean field (leading order $1/N$ or Hartree) 
approximations will be referred to as leading order (LO). It should also 
be noted that due to the effective resummation inherent in the 2PI 
approach, infinite series of perturbative diagrams are included.}. The 
reason for this situation is clear: beyond NLO the complexity of the 
equations and the numerical effort required to solve them increases 
dramatically \cite{Aarts:2006cv}. In this paper we present the first 
results beyond NLO in field theory.

Remarkably, the lowest order truncations beyond mean field theory include 
already many of the physical processes necessary to describe quantum field 
dynamics both far and close to equilibrium and are capable of capturing 
effective memory loss, universality of late time evolution, and 
thermalization. The natural question to ask is then how accurate a 
truncation at a given order describes the dynamics in the full theory. 
There are several ways this can be investigated.
 In the case of a systematic expansion, it should be possible to compare 
different orders of the expansion, shedding light on the effective 
convergence. When restricting to LO and NLO truncations only, the 
applicability of this approach is limited. The reason is that in LO mean 
field approximations scattering is absent and there is no notion of 
equilibration and thermalization, as there is at NLO. It is therefore 
necessary to consider the next-to-next-to-leading order (\NNLO) 
contribution as well.
 In the first part of this paper we study this problem in the $O(N)$ model 
and compare the dynamics obtained at LO, NLO and (part of) \NNLOs in the 
$1/N$ expansion in quantum field theory.
 In cases where an exact solution is available, a direct comparison can be 
carried out, and this approach has been successfully applied using 
classical statistical field theory instead of quantum field theory 
\cite{Aarts:2000wi,Aarts:2001yn}. In the classical limit, the 
nonperturbative solution can be constructed numerically by direct 
integration of the field equations of motion, sampling initial conditions 
from a given probability distribution. We use this approach to further 
quantify the role of truncations in the second part of the paper.

{\em 2PI dynamics --} 
 We consider a real \mbox{$N$-component} scalar quantum field with a 
$\lambda(\phi_a\phi_a)^2/(4! N)$ interaction ($a=1,\ldots,N$). In order to 
allow for an economical $1/N$ expansion, an auxiliary field $\chi$ is 
introduced to split the four-point interaction, which results in the 
action
 \bea
 S[\phi,\chi] = - \int_\C dx^0\int d\vecx \,\Big[ 
\half\partial_\mu\phi_a\partial^\mu \phi_a  
+ \half m^2 \phi_a\phi_a 
&&\nn \\ 
-\frac{3N}{2\lambda}\chi^2 + \half\chi\phi_a\phi_a 
\Big]. &&
\eea
 The theory is formulated along the Schwinger-Keldysh contour $\C$, as is
appropriate for initial value problems.
 In the symmetric phase ($\bra\phi_a\ket=0$), the 2PI effective action 
depends on the one-point function $\bar\chi=\bra\chi\ket$ and the 
two-point functions
\bea
G_{ab}(x,y) &=&  G(x,y) \delta_{ab} = 
\bra T_\C \phi_a(x) \phi_b(y)\ket, \\
D(x,y) &=&  \bra T_\C \chi(x) \chi(y)\ket -
\bra\chi(x)\ket\bra\chi(y)\ket.
\eea
In terms of those, the action is parametrized as 
\cite{Cornwall:1974vz,Aarts:2002dj}
\bea
\nn
 && \!\!\!\!\!\Gamma[G,D, \bar\chi] =
 S[0,\bar\chi]+ \frac{i}{2} \Tr\ln G^{-1} + \frac{i}{2} \Tr\, G_0^{-1}
\left(G-G_0\right) 
\\  &&  \;\;\;\;	
+ \frac{i}{2} \Tr\ln D^{-1} + \frac{i}{2} \Tr\, D_0^{-1} 
\left(D-D_0\right) + \Gamma_2[G,D],  
\label{eqGamma}
\eea
 where $G_0^{-1}=i(\square+m^2+\bar\chi)$ and $D_0^{-1}=3N/(i\lambda)$ 
denote the free inverse propagators. Extremizing the effective action 
gives equations for $\bar\chi$, $G$ and $D$. The latter take the 
standard form $G^{-1}=G_0^{-1}-\Sigma$ and $D^{-1}=D_0^{-1}-\Pi$, with 
self energies $\Sigma = 2i\delta \Gamma_2/\delta G$ and $\Pi = 2i\delta 
\Gamma_2/\delta D$.

In order to solve an initial value problem, the propagators and self 
energies are decomposed in statistical ($F$) and spectral ($\rho$) 
components. For instance, the basic two-point function is written as 
$G(x,y) = F(x,y)- (i/2){\rm sign}_{\C}(x^0-y^0) \rho(x,y)$, and
the causal equations for $F$ and $\rho$ take the form
 \bea
\nn
\left[\square_x +M^2(x)\right]F(x,y) =
- \int_0^{x^0}\!\!\! dz\, \Sigma_{\rho}(x,z)F(z,y) && \\
+ \int_0^{y^0}\!\!\! dz\, \Sigma_{F}(x,z) \rho(z,y), &&
\label{eqF}
\\
\left[\square_x +M^2(x)\right]\rho(x,y) =
-\int_{y^0}^{x^0}\!\!\! dz\, \Sigma_{\rho}(x,z)\rho(z,y). &&
\nn
\eea
 Here we used the notation $\int_{y^0}^{x^0}dz = \int_{y^0}^{x^0}dz^0\int 
d\vecz$. The effective mass parameter is given by 
 $M^2(x)=m^2+\bar\chi = m^2 +\lambda(N+2)/(6N)F(x,x)$. 
 After decomposing $D(x,y) = \lambda/(3N)\big[ i\delta_{\C}(x-y)+ \hat 
D_F(x,y) - (i/2){\rm sign}_\C(x^0-y^0) \hat D_\rho(x,y) \big]$, similar 
equations can be found for $\hat D_F$ and $\hat D_\rho$ 
\cite{Aarts:2002dj}.

\begin{figure}[t]
\centerline{\psfig{figure=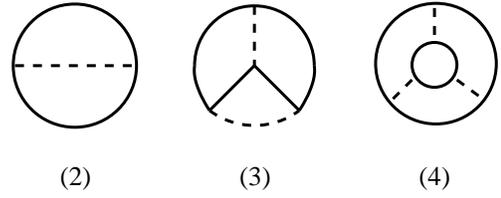,width=6.5cm}}
\caption{2PI$-1/N$ expansion: NLO (2 loops) and \NNLOs (3 and
4 loops) contributions. The full/dashed line denotes the $G/D$ 
propagator.
}
\label{fig1}
\end{figure}

{\em 2PI-$1/N$ expansion --}
 In the 2PI-$1/N$ expansion the nontrivial contribution $\Gamma_2[G,D]$ is 
written as $\Gamma_2 = \Gamma_2^{\rm NLO} + \Gamma_2^{\rm N^2LO} +\ldots$ 
Powercounting is straightforward: a closed loop of $G$ propagators yields 
a factor $N$, whereas a $D$ propagator contributes $1/N$. This yields one 
diagram at NLO ($\sim N^0$) and two diagrams at \NNLOs ($\sim 1/N$), see 
Fig.\ \ref{fig1}. Cutting a $G/D$ line gives the self energy $\Sigma/\Pi$. 
At NLO the self energies have no internal vertices and are therefore 
easily evaluated in real space, where they are given by the product of two 
propagators. For example, the statistical component of the NLO self energy 
is given by the expression
 \be
\Sigma_F^{\rm NLO}(x,y) =  
 -g \left[
 F(x,y)\hat D_F(x,y) - \frac{1}{4} \rho(x,y)\hat D_\rho(x,y) \right],
\label{eqNLO}
\ee
where $g=\lambda/(3N)$. 
 At \NNLO, the number of terms increases substantially, due to the 
possibility that any line can be of the $F$ or $\rho$ type. Moreover, self 
energies have two or four internal vertices, greatly increasing 
the complexity of the expressions. Let us first consider the 3-loop 
diagram at \NNLO. The corresponding self energy has two internal vertices 
and five propagators. The complete expression can be found in Ref.\ 
\cite{Aarts:2006cv} and has ${\cal O}(30)$ terms. Here we give two terms 
to illustrate the structure,
 \bea
&& \!\!\!\!\!\! \Sigma_{F}^{\rm N^2LO}(x,y) = 
- g^2 
\int_0^{x^0}\!\!\! dz \int_{z^0}^{y^0}\!\!\! dw\,  
\rho(x,z) \hat D_F(x,w)  \rho(y,w)  \nn \\ 
&& \hspace*{0.8cm}
\left[ \hat D_F(y,z) F(z,w) + \frac{1}{4} \hat D_\rho(y,z) \rho(z,w) 
\right] + \ldots
\label{eqNNLO}
\eea
The two internal vertices result in two nested memory integrals. 
 The second diagram at \NNLOs yields self energies with four nested memory 
integrals. In contrast to NLO, at \NNLOs the evaluation of self energies 
completely dominates the numerical effort.

{\em Quantum dynamics far from equilibrium --}
 We consider a spatially homogeneous system and solve the dynamical 
equations numerically in $1+1$ dimensions, by discretizing the system on a 
lattice with spatial lattice spacing $a$ and temporal lattice spacing 
$a_t$, using $a_t/a=0.2$. Initial conditions are determined by a Gaussian 
density matrix, resulting in initial correlation functions $F(0,0;\kv) = 
F(a_t,a_t;\kv) = [ n_0(\om_\kv)+1/2]/\om_\kv$, $F(a_t,0;\kv) = 
F(0,0;\kv)(1-a_t^2\om_\kv^2/2)$. Here $n_0$ is the initial 
(nonequilibrium) particle number and 
$\om_\kv=\left(\kv^2+M^2_0\right)^{1/2}$ with $M_0$ the initial mass, 
determined selfconsistently from the mean field mass gap equation. Initial 
conditions for the spectral function are determined by the equal time 
commutation relations.

We consider LO mean field dynamics, obtained by putting the nonlocal 
memory integrals on the right-hand side of Eq.\ (\ref{eqF}) to zero. The 
only nonequilibrium aspect is in the time dependent mass $M^2(x)$. The NLO 
approximation is solved without further approximation. At \NNLOs the 
numerical effort differs substantially between the 3-loop and the 4-loop 
diagram in Fig.\ \ref{fig1}, due to 2 and 4 internal nested memory 
integrals respectively. Therefore we restrict ourselves to the first 
diagram at \NNLOs and refer to this as \NNLO$'$. 
 In principle, a subtle cancellation between the 3- and 4-loop diagram 
could occur. However, since the second \NNLOs diagram is (naively) 
suppressed by two powers of the coupling constant, this is found not to 
be the case. We come back to this below. 
 A similar comparison between 2- and 3-loop truncations, appearing at 
the same order in a coupling expansion, was made in Ref.\ 
\cite{Arrizabalaga:2005tf}.

\begin{figure}[t] 
\centerline{\epsfig{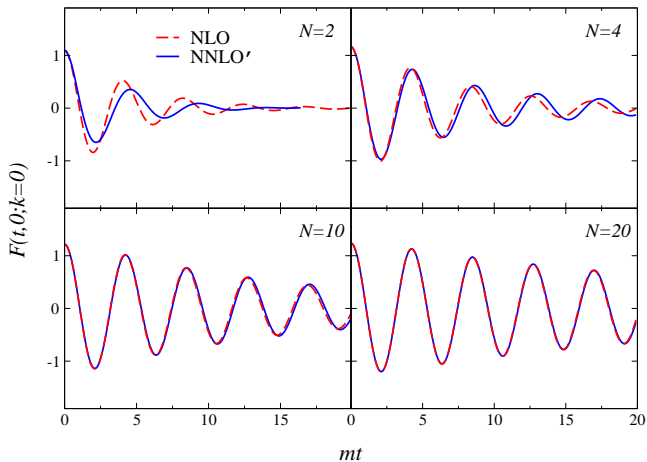}} 
\caption{ 
 Unequal time correlation function $F(t,0;\kv={\bf 0})$ 
for $N=2, 4, 10, 20$ in the quantum theory. The dashed/full lines 
show results for the 2PI-$1/N$ expansion at NLO/\NNLO$'$.
 }
\label{fig2} 
\end{figure} 

\begin{figure}[t] 
\centerline{\epsfig{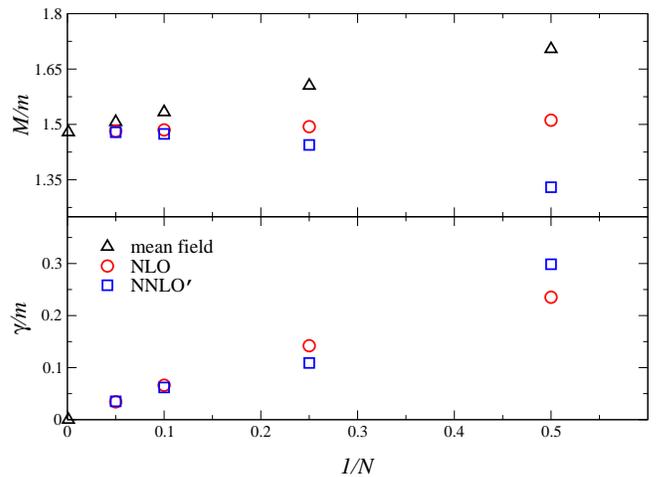}} 
\caption{ 
 Masses and widths extracted from the unequal-time correlation functions 
in Fig.\ \ref{fig2} as a function of $1/N$. In the mean field
approximation, the width is zero for all values of~$N$.
 } 
\label{fig3} 
\end{figure}

The unequal-time two-point function at zero momentum is shown in Fig.\ 
\ref{fig2}. The coupling is $\lambda/m^2=30$, where $m$ is the 
renormalized mass in vacuum. The lattice spacing is $am=0.5$ and volume 
$Lm=32$. The memory kernel is preserved completely. At LO (not shown) 
there is no damping in the unequal-time correlation function. Beyond LO, 
we observe an underdamped oscillation, the signal of a well-defined 
quasiparticle. Increasing $N$ shows convergence of the two truncations 
considered. We can extract the quasiparticle properties far from 
equilibrium by fitting these curves to an Ansatz of the form $A e^{-\gamma 
t}\cos Mt$, where $M$ is the quasiparticle mass and $\gamma$ its width 
(divided by 2). The resulting values are shown in Fig.\ \ref{fig3} as a 
function of $1/N$. We observe effective convergence when the value of $N$ 
is increased, as expected for a controlled expansion. For $N\gtrsim 10$, 
the NLO and \NNLO$'$ results are practically indistinguishable. In $1+1$ 
dimensions the perturbative onshell width from $2\to 2$ scattering 
processes is not defined due to the constraints from energy-momentum 
conservation. However, including higher order effects via the use of 
selfconsistently determined propagators leads to a width that vanishes as 
$1/N$, as expected from naive powercounting. The quasiparticle mass 
converges to the value determined by the nonthermal fixed point of the 
mean field equations in the large $N$ limit, which can be computed 
analytically \cite{Aarts:2000wi}. For the parameters used here we find 
$M/m=1.48$ when $N\to \infty$.

{\em Classical statistical field theory --}
 In order to further investigate the applicability of the $1/N$ expansion 
and in particular the role of the 4-loop diagram at \NNLO, we turn to 
classical statistical field theory, where the nonperturbative evolution 
can be computed by numerically solving the classical field equations, 
sampling initial conditions from the Gaussian ensemble. In order to carry 
out the comparison, we also take the classical limit in the set of 2PI 
equations. In this limit statistical ($F$) two-point functions dominate 
with respect to spectral ($\rho$) functions. There are several ways this 
can be motivated, using for instance classical statistical diagrams or 
arguing for classicality when occupation numbers are large 
\cite{Aarts:1997kp,Aarts:2001yn,Cooper:2001bd}. The result is 
that terms are dropped that are subleading when $F$ is taken to be much 
larger than $\rho$. For example, both in Eq.\ (\ref{eqNLO}) at NLO and in 
Eq.\ (\ref{eqNNLO}) at \NNLO, the last terms are dropped with respect to 
the first ones. We have carried out this procedure for all terms appearing 
in self energies at \NNLO$'$  in Ref.\ \cite{Aarts:2006cv}.

\begin{figure}[t] 
\centerline{\epsfig{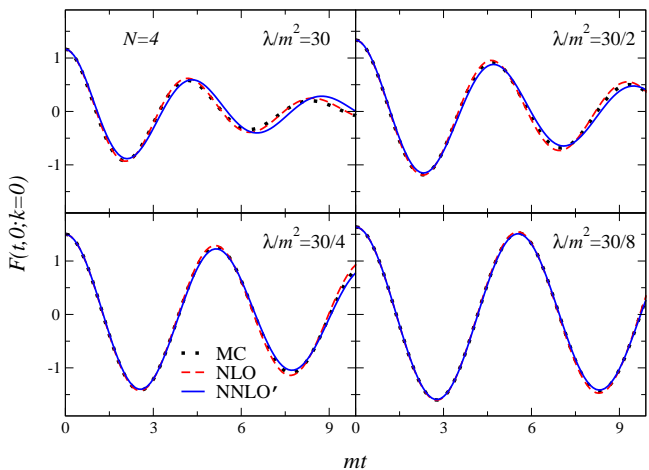}} 
\caption{
 Unequal time correlation function $F(t,0;\kv={\bf 0})$ for $N=4$ in the 
classical theory. Included are the NLO and \NNLO$'$ results in the 
classical limit of the 2PI-$1/N$ expansion, and the exact result obtained 
by numerical integration of the classical equations of motion (MC). The 
coupling constant is $\lambda/m^2=30/n$, with $n=1,2,4,8$.
 }
\label{fig5} 
\end{figure}

Since the 4-loop diagram at \NNLOs is suppressed by the coupling constant, 
its effect is expected to diminish at weaker coupling. In order to verify 
this, we use a relatively small value of $N=4$, where NLO and \NNLO$'$ 
differ at large coupling, both in the quantum and the classical theory.
 A comparison between NLO, \NNLO$'$, and the exact numerical result (MC 
for Monte Carlo) is shown in Fig.\ \ref{fig5}. In the numerical 
integration of the classical equations of motion, we have used a sample of 
$2.5\times 10^5$ initial conditions. In the top left corner, the coupling 
constant is $\lambda/m^2=30$. Damping at \NNLO$'$ is less than at NLO, 
similar to the situation in the quantum theory. Decreasing the coupling 
constant, we observe that the agreement between the two 2PI truncations 
improves. Moreover, for weaker coupling, we note that the MC results lie 
between the curves at NLO and \NNLO$'$, signalling a well converging 
expansion. For even smaller coupling (not shown), we found that all curves 
lie on top of each other, even for $N=4$.
 Since the \NNLOs approximation is numerically substantially more 
intensive than the NLO one, we have investigated whether the memory kernel 
at \NNLOs could be truncated earlier than at NLO, saving considerable 
computer resources. However, we found this not to be the case: in the 
results shown here it was necessary to preserve memory kernels over the 
entire history, also for the \NNLOs contribution. At this moment this 
prevents us from going to larger times and studying the approach to 
equilibrium beyond NLO.

{\em Summary --} 
 We have investigated convergence properties of the 2PI-$1/N$ expansion 
for nonequilibrium quantum fields. We included contributions at \NNLO, and 
found that the expansion is convergent even at large coupling as $N$ is 
increased. At fixed $N$, we found convergence as the coupling strength is 
reduced.
 These results confirm that the 2PI-$1/N$ expansion truncated at NLO 
already gives quantitatively accurate results for moderate values of $N$ 
or coupling strength: At any coupling $N=10$ works very well, whereas for 
the often used $N=4$, a trade-off in terms of a smaller coupling is 
required for precision calculations. We expect this behaviour to persist 
in $3+1$ dimensions, but a direct confirmation would be welcome.


\begin{acknowledgments}

The numerical work was conducted on the Murska Cluster at the Finnish 
Center for Computational Sciences CSC. G.A.\ and N.L.\ are supported by 
STFC. A.T.\ is supported by Academy of Finland Grant 114371.

\end{acknowledgments}


\end{document}